\documentclass[%
reprint,
superscriptaddress,
amsmath,amssymb,
aps,
prl,
]{revtex4-2} 

\usepackage{graphicx}
\usepackage{dcolumn}
\usepackage{bm}
\usepackage{hyperref}
\usepackage[percent]{overpic}

\graphicspath{{figures/}}

\begin{document}

\title{Mitigation of the onset of hosing in the linear regime through plasma frequency detuning}

\author{Mariana Moreira}
 \email{mariana.t.moreira@tecnico.ulisboa.pt}
 \affiliation{%
 GoLP/Instituto de Plasmas e Fus\~ao Nuclear, Instituto Superior T\'ecnico, Universidade de Lisboa, 1049-001 Lisboa, Portugal
}%

\author{Patric Muggli}
\affiliation{Max Planck Institute for Physics, D-80805 Munich, Germany}%
\affiliation{CERN, CH-1211 Geneva, Switzerland}%

\author{Jorge Vieira}
\email{jorge.vieira@tecnico.ulisboa.pt}
\affiliation{%
 GoLP/Instituto de Plasmas e Fus\~ao Nuclear, Instituto Superior T\'ecnico, Universidade de Lisboa, 1049-001 Lisboa, Portugal
}%

\begin{abstract}

The hosing instability poses a feasibility risk for plasma-based accelerator concepts. We show that the growth rate for beam hosing in the linear regime (which is relevant for concepts that use a long driver) is a function of the centroid perturbation wavelength. We demonstrate how this property can be used to damp centroid oscillations by detuning the plasma response sufficiently early in the development of the instability. We also develop a new theoretical model for the early evolution of hosing. These findings have implications for the general control of an instability's growth rate.

\end{abstract}

\maketitle

In the last decades, plasma-based acceleration has been explored as a possible avenue for the next generation of particle accelerators. Plasma can sustain electric fields about a thousand times higher than conventional accelerator elements. This estimate is based on a plasma electron density $n_0 = 10^{18}~\mathrm{cm}^{-3}$ and the cold nonrelativistic wavebreaking field~\cite{e0_akhiezer,e0_dawson} $E_0 = m_e c \, \omega_p/e \, \approx \, 96 \sqrt{n_0 \left[ \mathrm{cm}^{-3} \right]} \, \left[ \mathrm{V/m} \right]$, where $c$ is the speed of light, $m_e$ is the electron mass, $e$ is the elementary charge, $\omega_p = \sqrt{e^2 n_0 / \varepsilon_0 \, m_e}$ is the electron plasma frequency, and $\varepsilon_0$ is the vacuum permittivity.

One approach to leveraging plasma-based acceleration for high-energy physics applications is to drive a plasma wave (and the associated wakefields) with a high-energy particle bunch and accelerate a trailing bunch in a single, many-meter-long plasma, thus avoiding the complexity of staging schemes~\cite{staging_lasers,staging_beams}. This concept has been demonstrated recently in the AWAKE experiment at CERN using a 400\nobreakdash-GeV proton bunch as the driver through a ten-meter-long plasma, leading to the acceleration of injected electrons from 18~MeV up to 2~GeV~\cite{awake_nat}.

A plasma wave is excited effectively by an impulse with a length of the order of a plasma wavelength $\lambda_p = 2 \pi c/ \omega_p$. It is currently impractical to compress a 400\nobreakdash-GeV proton bunch into this longitudinal size, even for lower plasma densities on the order of $10^{14}~\mathrm{cm}^{-3}$. Nevertheless, high-amplitude wakefields can be reached by letting the bunch undergo the self-modulation instability (SMI)~\cite{smi_kumar, schrsmi} and controlling its onset with seeding~\cite{awake_fabian,awake-eseeding}. The same transverse wakefields that modulate the beam envelope at $\sim \lambda_p$ (self-modulation) can also modulate the centroid of the bunch when it is not propagating axisymmetrically, leading to the hosing instability (HI)~\cite{whittum}. This oscillation of the drive bunch centroid can alter the structure of the wakefields significantly, thus impacting their quality and that of an accelerated witness bunch.

The long-bunch HI and the SMI have comparable growth rates and can couple to each other when growing from similar seed levels~\cite{schrcoupled,schrcoherent}. Nevertheless, hosing can be avoided in a fully self-modulated bunch. This requires strongly seeding the SMI~\cite{hosevieira}, which has been accomplished in experiments by letting an ionizing laser pulse propagate with the proton bunch and create the plasma~\cite{awake_fabian}. In the future, SMI seeding may be achieved with a preceding short electron bunch, such that the entire proton bunch self-modulates~\cite{awake-eseeding,awakeplans_patric,awakeplans_edda}. When misaligned, this arrangement may seed the growth of hosing.
For short bunches in both quasilinear and nonlinear wakefield regimes, it has been established that hosing eventually reaches saturation~\cite{timon,hoselehe}, leading to an increased emittance.
However, saturation in the long-bunch case is not yet well understood, both in the presence and absence of the SMI.
Since this could represent a major impediment for high-quality acceleration, it would be useful to develop further mitigation methods for hosing in the linear wakefield regime.

In this Letter we show that the growth rate at the onset of hosing in the linear regime of plasma wakefield acceleration is highly sensitive to the centroid perturbation wavelength, and we exploit this property to suppress the growth of the HI. As a proof of concept of this mitigation approach, we consider the front portion of a long, cold electron bunch propagating through several plasma density steps.
Nevertheless, these HI properties are general and their application as a mitigation strategy can be scaled for any relativistic particle driver through the betatron period.
We also develop a theoretical model that can describe the early evolution of hosing and supplement existing asymptotic models. Our work opens up an unprecedented avenue for the control of hosing, which is vital for the future of plasma-based acceleration.

We consider a particle bunch propagating along $z$ with a density profile $n_b(\zeta, y) = n_{b0} \: f(\zeta) \: g(y, y_c)$, where $n_{b0}$ is the peak density and $f$ and $g$ are the normalized longitudinal and transverse profiles, respectively. The longitudinal coordinate $\zeta = z - c t$ moves at the beam velocity $v_b \approx c$, while $y$ is the transverse coordinate and $y_c(\zeta,z) = \left< y \right>$ is the bunch centroid, where the angle brackets denote a bunch-weighted transverse average defined as $\left< \cdot \right> = \int \cdot \; n_b(\zeta,y) dy \: /\int n_b(\zeta,y) dy$. An equation to describe hosing can be derived from the transverse averaging of the motion of a single relativistic bunch particle:
\begin{equation} \label{eq:ycfy}
    \frac{d^2 y_c}{dz^2} = \frac{m_e}{\gamma M_b} \frac{\left< F_y \right>}{e E_0}  \quad ,
\end{equation}
\noindent where $M_b$ is the bunch particle mass and $F_y$ is the transverse component of the force associated with the plasma wakefields driven by the bunch.
All distances are normalized to the plasma skin depth $k_p^{-1} = c/\omega_p$ in Eq.~\eqref{eq:ycfy} and in the rest of this Letter.
Note that, for small centroid displacements ($y_c \ll 1$), the Green's function solution to the plasma operator contained in $\left< F_y \right>$ of Eq.~\eqref{eq:ycfy} can be Taylor-expanded into a linear function of $y_c$. This means that, in this approximation, Eq.~\eqref{eq:ycfy} can always be written in the form of a harmonic oscillator (HO) equation:
\begin{equation} \label{eq:ho}
\left( \frac{d^2}{dz^2} + k_\mathrm{HO}^2(\zeta,z) \right) y_c(\zeta,z) = F(\zeta,z, y_c) \quad ,
\end{equation}
\noindent where the specific forms of the natural wavenumber $k_\mathrm{HO}^2(\zeta,z)$ and the driving force $F(\zeta,z, y_c)$ depend on the choice of geometry and transverse profile of the bunch. We also note that a centroid velocity $v_c = c \: dy_c/dz$ may be defined in association with the development of hosing.

For a Gaussian transverse profile, and assuming a constant beam envelope $\sigma_y$, $\left< F_y \right>$ in 2D Cartesian geometry is given by~\cite{supmat}: \nocite{erfintegrals}
\begin{eqnarray} \label{eq:avefy}
    & \displaystyle \frac{\left< F_y \right>}{e E_0} = \sqrt{\frac{\pi}{8}} \frac{n_{b0}}{n_0} \left( \frac{q_b}{e} \right)^2  \sigma_y \: \exp(\sigma_y^2)   \int_\zeta^\infty d\zeta'  \sin( \zeta - \zeta' ) \: f(\zeta') \nonumber  \\*
    & \displaystyle \times \left\{ \exp\left[ y_c(\zeta')-y_c(\zeta) \right] \: \mathrm{erfc}\left[ \frac{y_c(\zeta')-y_c(\zeta)+2 \: \sigma_y^2}{2 \: \sigma_y} \right] \nonumber \right.  \\*
   & \displaystyle \left. -  \exp\left[y_c(\zeta)-y_c(\zeta') \right] \: \mathrm{erfc}\left[ \frac{y_c(\zeta)-y_c(\zeta')+2 \: \sigma_y^2}{2 \: \sigma_y} \right] \right\} \; ,
\end{eqnarray}
\noindent where $q_b$ is the bunch particle charge, and $\mathrm{erfc}(x)$ is the complementary error function.

For the following, we consider an initial perturbation of the bunch centroid of the form $y_{c0}(\zeta) \propto \sin(\hat{k} \: \zeta)$, where $\hat{k} = k/k_p$ is an arbitrary wavenumber. In order to measure the HI growth rate, we define the growth factor
\begin{equation}
    \Gamma(z) = \frac{\int_L |y_c(z,\zeta)| \: d\zeta}{\int_L |y_c(0,\zeta)| \: d\zeta } \quad ,
\end{equation}
\noindent where $L$ is the length of some region of interest along the bunch.
Here we will consider a window measuring $L = 140~k_p^{-1}$, which represents around 22~$\lambda_p$, and a bunch front at $\zeta_s = 135~k_p^{-1}$.

\begin{figure}
  \begin{overpic}[width=\linewidth]{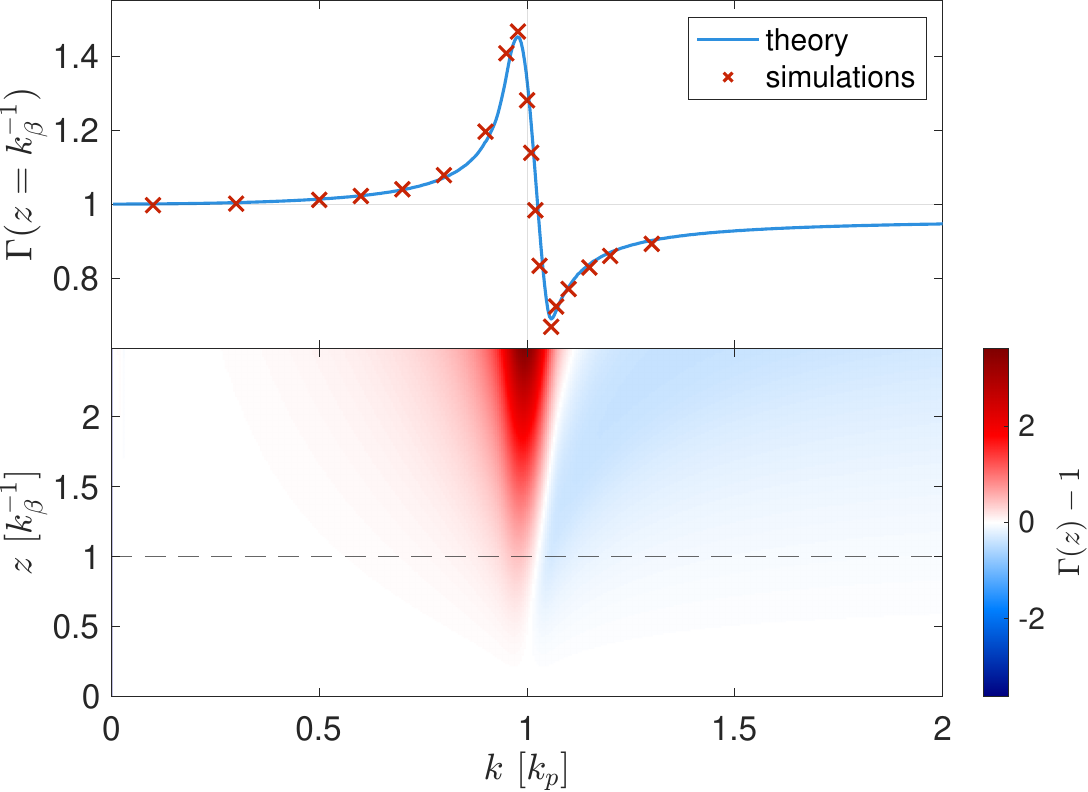}
    \put (13,68) {a)}
    \put (13,36) {b)}
  \end{overpic}
 \caption{\label{fig:gr}a) Hosing growth factor at $z = k_\beta^{-1}$ versus wavenumber of an initial sinusoidal perturbation, according to theory (blue curve) and simulations (red cross symbols). b) Theoretical hosing growth spectrum along $z$. The horizontal dashed line corresponds to the blue line-out in a).}
\end{figure}

During the initial propagation in plasma, we can assume a constant ($z$-independent) average transverse force $\left< F_{y0} \right>$ caused by $y_{c0}$ [replacing $y_c$ with $y_{c0}$ in Eq.~\eqref{eq:avefy}].
This assumption is valid within the typical timescale for bunch evolution, given by the betatron period $k_\beta^{-1} = c / \omega_\beta$ (defined by $\omega_\beta = \omega_b / \sqrt{2 \gamma}$, where $\omega_b = \sqrt{ n_{b0} q_b^2 / \varepsilon_0 M_b}$).
In this case the solution to Eq.~\eqref{eq:ycfy} is simply $y_c(\zeta, z) = y_{c0}(\zeta) + \tfrac{1}{2} z^2 \left( \frac{m_e}{\gamma M_b} \: \frac{\left< F_{y0} \right>}{e E_0} \right) $. To evaluate $\Gamma(z)$, we consider a relativistic ($\gamma = 480$) electron bunch with $n_{b0} = 0.001 \: n_0$, a Gaussian transverse profile with the rms transverse size $\sigma_y \approx 0.27 \: k_p^{-1}$, and a longitudinal profile $f(\zeta) = \frac{1}{2} \left(1 + \cos(\sqrt{\frac{\pi}{2}} \frac{\zeta - \zeta_c}{\sigma_z}) \right)$, where $\sigma_z \approx 160 \: k_p^{-1}$, $\zeta_c = \zeta_s - \sqrt{2 \pi} \: \sigma_z$, and with the limits $\zeta = \pm \sqrt{2 \pi} \: \sigma_z + \zeta_c$. The resulting curve (calculated numerically) is shown in Fig.~\ref{fig:gr}a) (blue) for $z = k_\beta^{-1}$, which is approximately $0.74 ~ \mathrm{m}$ for the above electron bunch parameters and $n_0 = 0.5 \times 10^{14}~\mathrm{cm}^{-3}$.

The theoretical curve is confirmed by 2D Cartesian particle-in-cell simulations~\cite{supmat} with OSIRIS~\cite{osiris} [red cross symbols in Fig.~\ref{fig:gr}a), where each symbol represents a simulation initialized with a centroid perturbation at $k$]. An identical theoretical curve can be obtained in cylindrical coordinates (assuming $y_c \ll 1$), and in fact we use 3D simulations later in this Letter to explore the consequences of this result.

The hosing frequency response in Fig.~\ref{fig:gr}a) (note that $k$ may be interpreted as a frequency via $f = k c/2 \pi$) implies that there is a finite growth rate for initial perturbations at frequencies other than the plasma frequency. The tail in the growth rate spectrum for $k < k_p$ has been discussed before in the context of long laser pulses under the term ``long-wavelength hosing''~\cite{longwvl1,longwvl2}. For short wavelengths ($k > k_p$), however, the growth factor is smaller than 1, i.e., there is a decrease of the initial perturbation amplitude instead of growth.
This property has not been previously predicted and underpins a new HI mitigation process that we explore in this Letter.
A further observation is that the maximum growth is not attained for $k = k_p$ (as generally assumed for the HI) during this initial phase of propagation, but at a slightly lower value ($k \approx 0.98~k_p$).

After some propagation in plasma, growth around $k_p$ becomes dominant, as illustrated in Fig.~\ref{fig:gr}b), based on the theoretical $\left< F_{y0} \right>$. Not only does the amplitude of resonant growth become several times that of negative growth, but the locations of both extrema, $k_\mathrm{max}$ and $k_\mathrm{min}$, shift with increasing $z$ ($k_\mathrm{max} \rightarrow k_p$ and $k_\mathrm{min} \rightarrow \infty$).
The typical duration of the transient phase where the maximum positive and negative growth amplitudes are similar [Fig.~\ref{fig:gr}a)] depends on the bunch properties, though in general this corresponds roughly to a propagation distance $ z \lesssim k_\beta^{-1}$.

\begin{figure}
  \begin{overpic}[width=\linewidth]{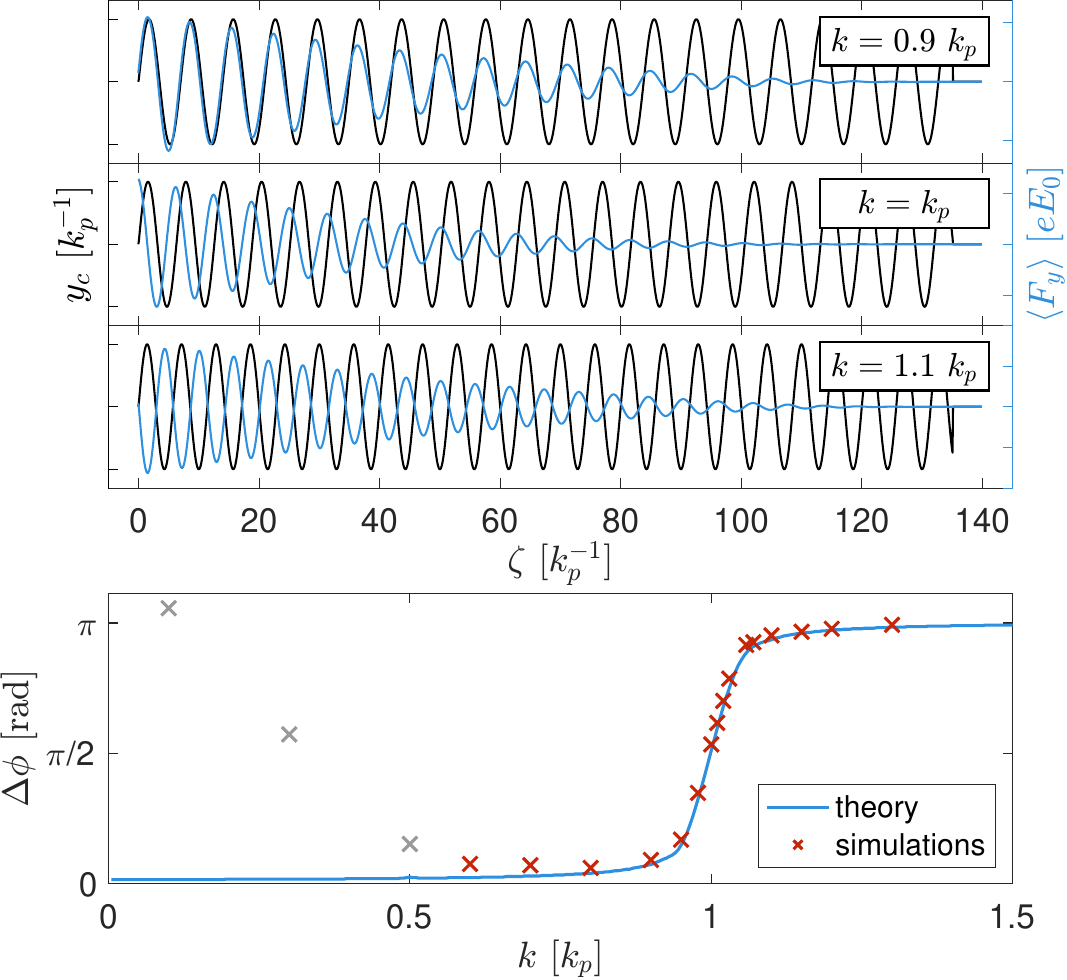}
    \put (0,88) {a)}
    \put (0,33) {b)}
  \end{overpic}
 \caption{\label{fig:phase} a) Initial centroid (black) and average transverse force (blue) for three different seed wavenumbers, obtained from 2D OSIRIS simulations at $z = 0$. b) Phase shift between the initial $y_c$ and $\left< F_y \right>$ as a function of the perturbation wavenumber, obtained from theory (blue curve) and simulations (red cross symbols).}
\end{figure}

We find that the phase shift between the bunch centroid perturbation and the transverse plasma response, represented by $\left< F_y \right>$, determines the early HI growth dynamics in Fig.~\ref{fig:gr}a).
Figure~\ref{fig:phase}a) displays the initial centroid and $\left< F_y \right>$ from three simulations at different perturbation wavenumbers, thereby illustrating three regimes.
For $k = 0.9~k_p$ (slow growth), $\left< F_y \right>$ is almost in phase with $y_c$, while the damping regime results from the fully out-of-phase transverse force, which acts in the opposite direction of $y_c$ at every $\zeta$ [see Fig.~\ref{fig:phase}a), $k = 1.1 \: k_p$]. For $k = k_p$ (resonant growth), the wakefield response lags the centroid perturbation by $\pi/2$.

Since $\left< F_y \right>$ initially oscillates at $k$ (along $\zeta$), a phase shift $\Delta \phi$ between both periodic curves can be measured straightforwardly. The relationship between $\Delta \phi$ and $k$ is shown in Fig.~\ref{fig:phase}b), as obtained from the 2D simulations and the theoretical $\left< F_{y0} \right>$, using a cross-correlation method~\cite{supmat}.
There is excellent agreement between theory and simulations for $k > 0.5 \: k_p$.
The simulation data points for $k \leq 0.5 \: k_p$ [grey cross symbols in Fig.~\ref{fig:phase}b)] are not valid, since the simulation window length becomes comparable to the perturbation wavelength ($L/\lambda \lesssim 10 $). The theoretical curve was obtained by scaling the nominal window and bunch lengths by $1/k$, thus encompassing several periods in the analysis.
The three growth regimes are again evident: two asymptotes at $\Delta \phi = 0$ and $\Delta \phi = \pi$, and a transition region where $\Delta \phi$ crosses $\pi/2$.

We note that the physics described in this section is characteristic of a sinusoidally driven damped HO: consider the phase jump when the driving frequency crosses the resonance, or how the curve for $\Gamma(z)$ at large $z$ [see Fig.~\ref{fig:gr}b)] resembles the HO amplitude curve as a function of driving frequency.

We wish to predict the development of the centroid displacement and velocity during the initial propagation in plasma and for an arbitrary seed wavenumber. Previous asymptotic models are inappropriate for our purposes, since they only apply far enough behind the front ($k_p \zeta \gg k_\beta z$) and assume initial centroid fluctuations at $k_p$~\cite{schrcoupled}.

The early evolution of $y_c$ can be approximated by truncating a series solution to Eq.~\eqref{eq:ycfy}~\cite{chao}, which may be written as $\partial_z^2 y_c = \mathcal{R}\{ y_c \}$, with the operator $\mathcal{R}$ in the $\zeta$ variable. For this method to be tractable, $\mathcal{R}$ should be a linear function of $y_c$, which is always the case assuming $y_c \ll 1$. Writing $y_c$ as a power series, $y_c = \sum_{n=0}^\infty a_n z^n$, we arrive at the following general solutions for $y_c$ and $v_c$ (normalized to $k_p^{-1}$ and $c$, respectively): 
\begin{eqnarray}
        y_c(\zeta, z) = \sum_{n = 0}^\infty \frac{z^n}{n!} \begin{cases}
    \mathcal{R}^{(\frac{n}{2})} \{ y_{c0} \} \; ,  & n \text{ is even} \\
    \mathcal{R}^{(\frac{n-1}{2})} \{ v_{c0} \} \; , & n \text{ is odd}
    \end{cases} \label{eq:ycsol} \\
        v_c(\zeta, z) = \sum_{n = 0}^\infty \frac{z^n}{n!} \begin{cases}
    \mathcal{R}^{(\frac{n}{2})} \{ v_{c0} \} \; ,  & n \text{ is even} \\
    \mathcal{R}^{(\frac{n+1}{2})} \{ y_{c0} \} \; , & n \text{ is odd}
    \end{cases} \label{eq:vcsol}
\end{eqnarray}
\noindent where $y_{c0}(\zeta)$ and $v_{c0}(\zeta)$ are arbitrary initial conditions for the centroid and centroid velocity, respectively, and the superscript in the operator $\mathcal{R}$ signifies consecutive applications (e.g. $\mathcal{R}^{(2)}\{ X \} = \mathcal{R}\{ \mathcal{R} \{ X \} \}$ and $\mathcal{R}^{(0)}\{ X \} = X$). In cylindrical coordinates and for a transverse flat-top profile $g(r) = \Theta[ r_{b0} - r]$, where $r_{b0}$ is a constant bunch radius, $\mathcal{R}\{X(\zeta, z)\} = 2  \hat{k}_\beta^2 \: I_1(r_{b0}) K_1(r_{b0}) \int_\zeta^\infty d\zeta' \sin( \zeta - \zeta' ) \: f(\zeta') \left[ X(\zeta,z) - X(\zeta',z) \right]$~\cite{schrcoupled}.

\begin{figure}
  \begin{overpic}[width=\linewidth]{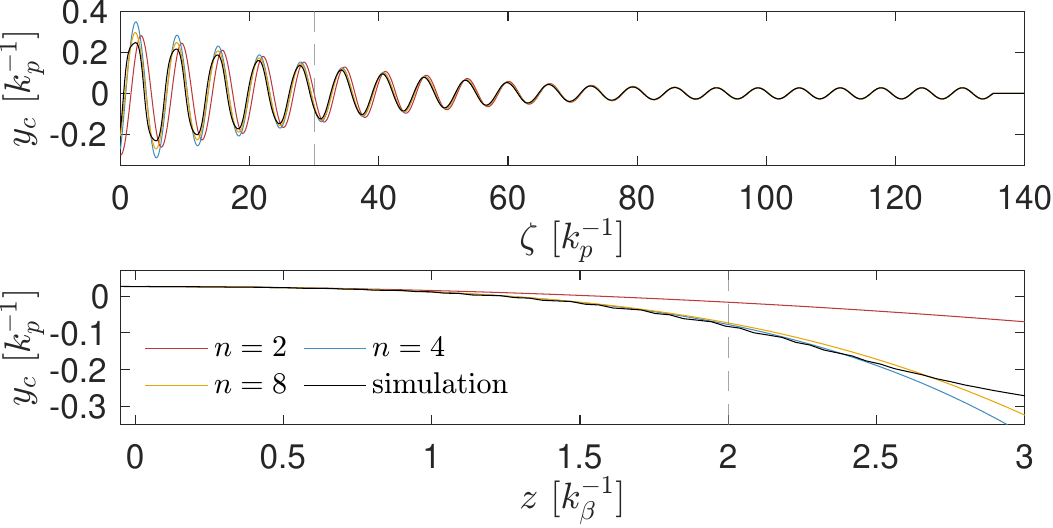}
    \put (92,45.5) {a)}
    \put (92,21) {b)}
  \end{overpic}
 \caption{\label{fig:evolmodel} Comparison of the series model truncated at $n$ terms (colored curves) with a 3D OSIRIS simulation (black curves), a) along $\zeta$ at $z = 2~k_\beta^{-1}$ and b) along $z$ at $\zeta = 30~k_p^{-1}$. }
\end{figure}

Figure~\ref{fig:evolmodel} compares the model to a 3D simulation of a transversely flat-top electron bunch with  $r_{b0} \approx 0.27 \: k_p^{-1}$ and an initial centroid $y_{c0}(\zeta) = 0.027 \sin(\zeta)$ propagating in constant-density plasma~\cite{supmat}. For the range in $\zeta$ we are considering, the truncated solutions Eq.~\eqref{eq:ycsol} and~\eqref{eq:vcsol} are valid up to propagation distances $z \sim k_\beta^{-1}$ [see Fig.~\ref{fig:evolmodel}b)]. A larger number of series terms improves the agreement between the model and simulation, as demonstrated in Fig.~\ref{fig:evolmodel}.

One could use the frequency response shown in Fig.~\ref{fig:gr}a) to, for example, damp the HI. In reality we may not control $k$, but we can control the local plasma density and therefore the ratio of $k$ (fixed in the centroid perturbation) to the local plasma wavenumber $k_p$, thereby operating in different growth regimes (e.g. at $\Gamma < 1$).
Initially, $y_c$ and $v_c$ react differently to each regime since they are phase-shifted by $\pi/2$~\cite{supmat}, and the disruption of the bunch can therefore only be minimised by alternating between regimes.
A mitigation set-up would therefore require propagating the hosed or seeded bunch through plasma sections with different densities. The series model can be applied to this set-up as long as we assume that the plasma density does not change at the length scale of $k_\beta^{-1}$. In that case we can normalize the entire model to the local plasma density.

\begin{figure}
 \includegraphics[width=\linewidth]{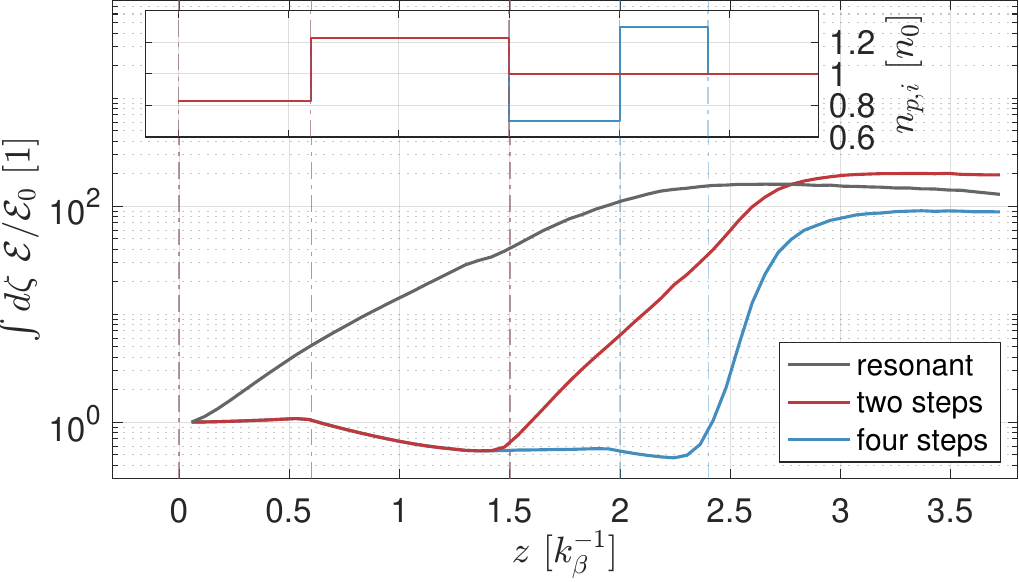}
 \caption{\label{fig:ene} Transverse energy of the bunch relative to the initial value along the plasma (integrated over the simulation domain): plasma at $n_0$ (grey), with two density steps (red), and with four density steps (blue). Results from 3D OSIRIS. Inset: plasma density profiles of the two density step configurations. The vertical dash-dotted lines indicate the boundaries of each density step.}
\end{figure}

Mitigation of a sinusoidal hosing seed at $k_p$ can be achieved by a series of plasma density steps with densities above and below $n_0$.
In fact, the use of a density step to control an instability has been proposed before (e.g.~\cite{lotov_step}).
Note that, for an arbitrarily shaped initial centroid, the fastest growing mode is sinusoidal, which means that this strategy can suppress the growing component of any seed. The two configurations chosen to demonstrate the concept are illustrated in Fig.~\ref{fig:ene} (inset). The parameters for these density steps were found by experimenting with the series model. Nevertheless, an exhaustive parameter scan could be achieved with machine-learning-based optimization algorithms.

We performed three 3D simulations (same parameters as before \cite{supmat}): two where a bunch seeded with $y_{c0}(\zeta)$ propagates through each of the two density profiles shown in Fig.~\ref{fig:ene} (inset) and in resonant plasma ($n_0$) afterwards, and one where the same bunch propagates exclusively in resonant plasma.

To quantify the effect of the density steps on the bunch and its hosing seed, we define a transverse energy $\mathcal{E}(\zeta,z) = \frac{1}{2} \: v_c^2(\zeta,z) +  \frac{1}{2} \: k_\mathrm{HO}^2(\zeta,z) \:  y_c^2(\zeta,z)$, in analogy with the HO. We note that the energy conservation equation for $\mathcal{E}(\zeta,z)$ can be obtained by multiplying Eq.~\eqref{eq:ho} by $v_c$. The wavenumber $k_\mathrm{HO}^2(\zeta,z)$ is computed directly from the $z$-evolving simulation data. In this geometry, $k_\mathrm{HO}^2(\zeta,z) = 2 \hat{k}_\beta^2 \: I_1[r_b(\zeta,z)]/r_b(\zeta,z) \int_\zeta^\infty d\zeta' \sin( \zeta - \zeta' ) \: f(\zeta',z) \; K_1[r_b(\zeta',z)] \; r_{b0}^2(\zeta',z)/r_b(\zeta',z)$, where 
$f(\zeta',z)$ is measured on axis.

Figure~\ref{fig:ene} compares $\mathcal{E}(\zeta,z)$ (integrated along $\zeta$ for a complete picture of the bunch at every time step and normalised to the initial value $\mathcal{E}_0$) for the three simulation cases. After two density steps (red curves), the energy is smaller than in the resonant case by almost two orders of magnitude (and lower than $\mathcal{E}_0$). It is possible to stack further pairs of density steps with tuned parameters so as to extend the suppressive effect on the HI, as demonstrated by the blue curves in Fig.~\ref{fig:ene}.

After the steps, however, the growth rate tends to be exacerbated and saturation is reached at similar energy values.
This exacerbation is at least partly explained by the self-focusing felt by the larger amount of charge around the axis (relative to the resonant case), and may be weakened by beam emittance and matching.
Regardless, placed before a separate acceleration stage, this set-up could conceivably suppress a hosing seed until a different mechanism (e.g. the SMI) dominates the beam dynamics.
Further work is necessary to explore different density profile configurations and their ability to control the growth and saturation of the HI.

In conclusion, we show that the growth rate of hosing in the linear regime depends on the frequency of the centroid perturbation, and that the transient growth spectrum at the onset of the instability can be harnessed to suppress its growth. 
We demonstrate how this could be achieved by detuning the background plasma frequency, i.e., by introducing a series of plasma density steps. These findings apply both to long particle bunches and laser pulses, and can be generalized for the control of both the hosing and self-modulation instability growth rates.
Additionally, the oscillator interpretation discussed here represents an alternative approach to understanding the onset of other beam-plasma instabilities.

\begin{acknowledgments}
M.M. thanks John Farmer for fruitful discussions and acknowledges the supervision of Bernhard Holzer.
This work was partially supported by the Portuguese Foundation for Science and Technology (FCT) through Grants No. CERN/FIS-TEC/0032/2017, No. CERN/FIS-TEC/0017/2019, and No. AWAKE-CERN/FIS-TEC/0034/2021; and by the European Union’s Framework Programme for Research and Innovation Horizon 2020 (2014--2020) under Grant Agreement No. 730871.
We acknowledge PRACE for awarding us access to Piz Daint at CSCS, Switzerland, and MareNostrum at Barcelona Supercomputing Center (BSC), Spain. We acknowledge the EuroHPC Joint Undertaking for awarding this project access to the EuroHPC supercomputer LUMI, hosted by CSC (Finland) and the LUMI consortium through a EuroHPC Regular Access call.
\end{acknowledgments}

\bibliography{references}

\end{document}


\title{Supplemental Material for \\ ``Mitigation of the onset of hosing in the overdense regime through plasma frequency detuning''}

\author{Mariana Moreira}
 \email{mariana.t.moreira@tecnico.ulisboa.pt}
  \affiliation{%
 GoLP/Instituto de Plasmas e Fus\~ao Nuclear, Instituto Superior T\'ecnico, Universidade de Lisboa, 1049-001 Lisboa, Portugal
}

\author{Patric Muggli}
\affiliation{Max Planck Institute for Physics, D-80805 Munich, Germany}%
\affiliation{CERN, CH-1211 Geneva, Switzerland}%

\author{Jorge Vieira}%
\affiliation{%
 GoLP/Instituto de Plasmas e Fus\~ao Nuclear, Instituto Superior T\'ecnico, Universidade de Lisboa, 1049-001 Lisboa, Portugal
}

\maketitle

\tableofcontents

\section{\label{sec:sm1} Derivation of the average transverse force for a Gaussian transverse beam profile in 2D Cartesian geometry}

We consider a particle bunch propagating along $z$ with a density profile $n_b(\zeta, y) = n_{b0} \cdot f(\zeta) \cdot g(\zeta,y)$, where $n_{b0}$ is the peak density and $f$ and $g$ are the normalized longitudinal and transverse profiles, respectively. The charge of the bunch particles is $q$. The longitudinal coordinate $\zeta = z - c \: t$ moves at the beam velocity $v_b \approx c$, while $y$ is the transverse coordinate and $y_c = \left< y \right>$ is the centroid, where the angle brackets denote a transverse average defined as $\left< \cdot \right> = \int \cdot~n_b(\zeta,y)~dy \: /\int n_b(\zeta,y)~dy$.

An equation to describe hosing can be derived from the transverse averaging of the motion of a single relativistic beam particle:
\begin{equation} \label{eq:ycfy}
    \frac{d^2 y_c}{dz^2} = \frac{m_e}{\gamma M_b} \frac{\left< F_y \right>}{e E_0}  \quad ,
\end{equation}
\noindent where all distances are normalized to the plasma skin depth $k_p^{-1} = c/\omega_{pe}$, $M_b$ is the beam particle mass, and $F_y$ is the transverse component of the force associated with the plasma wakefields driven by the beam. The transverse force is determined by calculating the wake potential $\psi$ and translating it into a force via $\frac{F_y}{e E_0} = - q/e \: \partial_y \psi $. Once we apply the transverse average, the latter equation becomes
\begin{equation} \label{eq:avefy}
    \frac{\left< F_y \right> }{e E_0} = - \frac{q}{e} \left< \partial_y \psi \right> \quad .
\end{equation}

A differential equation for the wake potential can be derived from the cold plasma fluid and Maxwell equations. In Cartesian geometry this equation is
\begin{equation}
    (\partial_\zeta^2 +1)(\partial_y^2-1) \: \psi = \frac{q}{e} \frac{n_b(\zeta,y)}{n_0} \quad ,
\end{equation}
\noindent for which the general Green's function solution is (see for example~\cite{schrcoherent})
\begin{equation} \label{eq:psicart}
    \psi = \frac{q}{e} \: \frac{n_{b0}}{n_0} \int_\zeta^\infty d\zeta' \sin(\zeta-\zeta') f(\zeta') \int_{-\infty}^{\infty} dy' \left( - \tfrac{1}{2} \right) \: e^{-y_>} \: e^{y_<} \: g(\zeta',y') \quad ,
\end{equation}
\noindent where $y_>$ and $y_<$ are, respectively, the larger and smaller of $y$ and $y'$. After replacing the wake potential solution (Eq.~\ref{eq:psicart}), Eq.~\ref{eq:avefy} may be written as
\begin{equation} \label{eq:avefytau}
    \frac{\left< F_y \right> }{e E_0} = - \left( \frac{q}{e} \right)^2 \frac{n_{b0}}{n_0} \int_\zeta^\infty d\zeta' \sin(\zeta-\zeta') f(\zeta') \: \left< \partial_y \tau(\zeta',y) \right> \quad ,
\end{equation}
\noindent where $\tau(\zeta,y)$ is defined as $\tau(\zeta,y) = - \tfrac{1}{2} \int_{-\infty}^{\infty} dy' e^{-y_>} \: e^{y_<}  \: g(y',\zeta)$. To obtain Eq.~3 of the article, we must simplify the expression $\left< \partial_y \tau(\zeta',y) \right>$ for a Gaussian transverse profile, i.e. $g(\zeta,y) = \exp\left( - \frac{\left(y-y_c(\zeta)\right)^2}{2 \: \sigma_y(\zeta)} \right)$, where $\sigma_y(\zeta)$ is the rms beam size in the $y$ direction.

Substituting the transverse profile in $\tau(\zeta,y)$, we obtain
\begin{equation}
\tau(\zeta,y) = - \frac{1}{2} \left[ e^{-y} \int_{-\infty}^y dy' \exp\left( y' - \frac{(y'-y_c(\zeta))^2}{2 \: \sigma_y^2(\zeta)} \right) + e^y \int_y^{\infty} dy' \exp\left(- y' - \frac{(y'-y_c(\zeta))^2}{2 \: \sigma_y^2(\zeta)} \right) \right] \; ,
\end{equation}
\noindent
which, after differentiation in $y$, leads to
\begin{equation}
\begin{split}
    \partial_y \tau(\zeta,y) =& - \frac{1}{2} \left[ - e^{-y} \int_{-\infty}^y dy' \exp\left( y' - \frac{(y'-y_c(\zeta))^2}{2 \: \sigma_y^2(\zeta)} \right) + e^y \int_y^{\infty} dy' \exp\left(- y' - \frac{(y'-y_c(\zeta))^2}{2 \: \sigma_y^2(\zeta)} \right) \right] \\
    =& - \frac{1}{2} \: \sqrt{\frac{\pi}{2}} ~ \sigma_y(\zeta) \exp\left( -y-y_c(\zeta)+\frac{\sigma_y^2(\zeta)}{2} \right) \left[ e^{2 y} ~ \mathrm{erfc}\left( \frac{y-y_c(\zeta)+\sigma_y^2(\zeta)}{\sqrt{2} \: \sigma_y(\zeta)} \right) \right. \\
    & \left. - e^{2 \: y_c(\zeta)} ~ \mathrm{erfc}\left( \frac{y_c(\zeta)-y+\sigma_y^2(\zeta)}{\sqrt{2} \: \sigma_y(\zeta)} \right) \right] \; ,
\end{split}
\end{equation}
\noindent where $\mathrm{erfc}(x) = 1-\mathrm{erf}(x)$ is the complementary error function. When we apply the transverse average to $\partial_y \tau(\zeta',y)$, where $\zeta'$ is the variable that will be integrated over in the right-hand side to Eq.~\ref{eq:avefytau}, we can simplify to
\begin{equation}
\begin{split}
   \left< \partial_y \tau(\zeta',y) \right> =& \frac{\int_{-\infty}^\infty dy ~ n_b(\zeta,y) ~ \partial_y \tau(\zeta',y) }{ \int_{-\infty}^\infty dy ~ n_b(\zeta,y) } = \frac{n_{b0} f(\zeta) \int_{-\infty}^\infty dy ~ g(\zeta,y) ~ \partial_y \tau(\zeta',y) }{ n_{b0} f(\zeta) \int_{-\infty}^\infty dy ~ g(\zeta,y) } \\
   =& \frac{1}{\sqrt{2 \pi} \: \sigma_y(\zeta)} \int_{-\infty}^\infty dy ~ g(\zeta,y) ~ \partial_y \tau(\zeta',y) \; .
\end{split}
\end{equation}
The resulting integral, after the transverse Gaussian profile is replaced again for $g(\zeta,y)$, is
\begin{equation} \label{eq:avedytau}
\begin{split}
    \left< \partial_y \tau(\zeta',y) \right> =& - \frac{1}{4} \frac{\sigma_y(\zeta')}{\sigma_y(\zeta)} \; e^{\sigma_y^2(\zeta')/2} \int_{-\infty}^{\infty} dy \; \exp\left[ - \frac{(y-y_c(\zeta))^2}{2 \; \sigma_y^2(\zeta)} \right] e^{-y-y_c(\zeta')} \left[ e^{2 y} \; \mathrm{erfc}\left( \frac{y-y_c(\zeta')+\sigma_y^2(\zeta')}{\sqrt{2} \; \sigma_y(\zeta')} \right) \right. \\
    &\left. - e^{2 \: y_c(\zeta')} \; \mathrm{erfc}\left( \frac{y_c(\zeta')-y+\sigma_y^2(\zeta')}{\sqrt{2} \; \sigma_y(\zeta')} \right)  \right] \quad ,
\end{split}
\end{equation}
\noindent and can be written as four terms [using $\mathrm{erfc}(x) = 1-\mathrm{erf}(x)$]. Two of these terms have the following solutions:
\begin{eqnarray}
    \begin{aligned}
        \exp\left( - y_c(\zeta') - \frac{y_c^2(\zeta)}{2 \sigma_y^2(\zeta)} \right) \int_{-\infty}^{\infty} dy \; \exp\left[ -\frac{y^2}{2 \sigma_y^2(\zeta)} + y \right. & \left. \left( 1+ \frac{y_c(\zeta)}{\sigma_y^2(\zeta)} \right) \right] = \\
        &\sqrt{2 \pi} ~\sigma_y(\zeta) ~\exp\left( y_c(\zeta) - y_c(\zeta') + \frac{\sigma_y^2(\zeta)}{2} \right) \; ,
    \end{aligned} \\
    \begin{aligned}
        - \exp\left( y_c(\zeta') - \frac{y_c^2(\zeta)}{2 \sigma_y^2(\zeta)} \right) \int_{-\infty}^{\infty} dy \; \exp\left[ -\frac{y^2}{2 \sigma_y^2(\zeta)} - y \right. & \left. \left( 1- \frac{y_c(\zeta)}{\sigma_y^2(\zeta)} \right) \right] = \\
        &- \sqrt{2 \pi} ~\sigma_y(\zeta) ~\exp\left( y_c(\zeta') - y_c(\zeta) + \frac{\sigma_y^2(\zeta)}{2} \right) \; .
    \end{aligned}
\end{eqnarray}
The other two terms can be solved using the following identity~\cite{erfintegrals}:
\begin{equation}
    \int_{-\infty}^\infty \exp\left( -a z^2 + \beta z \right) \mathrm{erf}\left( a_1 z + \beta_1 \right) dz = \sqrt{\frac{\pi}{a}} \exp\left( \frac{\beta^2}{4 a} \right)  \mathrm{erf}\left[ \frac{2 a \beta_1 + a_1 \beta}{2 \sqrt{a^2 + a a_1^2}} \right] \; , \quad a > 0 \: .
\end{equation}
The two remaining terms of the integral in Eq.~\ref{eq:avedytau} therefore become
\begin{eqnarray}
    \begin{aligned}
        \int_{-\infty}^\infty \exp \left[ - \frac{\left( y - y_c(\zeta) \right)^2}{2 \: \sigma_y^2(\zeta)} \right] & e^{-y-y_c(\zeta')}~e^{2 \: y_c(\zeta')}~\mathrm{erf}\left( \frac{y_c(\zeta') - y + \sigma_y^2(\zeta')}{\sqrt{2} \: \sigma_y(\zeta')} \right) dy = \\
        & \sqrt{2 \pi} ~\sigma_y(\zeta) ~\exp\left( y_c(\zeta') -y_c(\zeta) + \frac{\sigma_y^2(\zeta)}{2} \right) ~\mathrm{erf}\left[ \frac{ y_c(\zeta') -y_c(\zeta) +\sigma_y^2(\zeta) +\sigma_y^2(\zeta') }{\sqrt{2 \: \sigma_y^2(\zeta') - 2 \: \sigma_y^2(\zeta)}} \right] \: ,
    \end{aligned} \\
    \begin{aligned}
        - \int_{-\infty}^\infty \exp \left[ - \frac{\left( y - y_c(\zeta) \right)^2}{2 \: \sigma_y^2(\zeta)} \right] & e^{-y-y_c(\zeta')}~e^{2  y}~\mathrm{erf}\left( \frac{y - y_c(\zeta') + \sigma_y^2(\zeta')}{\sqrt{2} \: \sigma_y(\zeta')} \right) dy = \\
        & - \sqrt{2 \pi} ~\sigma_y(\zeta) ~\exp\left( y_c(\zeta) -y_c(\zeta') + \frac{\sigma_y^2(\zeta)}{2} \right) ~\mathrm{erf}\left[ \frac{ y_c(\zeta) -y_c(\zeta') +\sigma_y^2(\zeta) +\sigma_y^2(\zeta') }{\sqrt{2 \: \sigma_y^2(\zeta') + 2 \: \sigma_y^2(\zeta)}} \right] \: .
    \end{aligned}
\end{eqnarray}
Replacing the solutions to the four integrals in Eq.~\ref{eq:avedytau}, we obtain
\begin{equation} \label{eq:avedytausol}
    \begin{split}
     \left< \partial_y \tau(\zeta',y) \right> =& - \sqrt{\frac{\pi}{8}} \: \sigma_y(\zeta) \: \exp \left(\frac{\sigma_y^2(\zeta) + \sigma_y^2(\zeta')}{2} \right) \left\{ e^{ y_c(\zeta)-y_c(\zeta')} \: \mathrm{erfc}\left( \frac{y_c(\zeta)-y_c(\zeta')+\sigma_y^2(\zeta)+\sigma_y^2(\zeta')}{\sqrt{2 \: \sigma_y^2(\zeta) + 2 \: \sigma_y^2(\zeta')}} \right) \right. \\
    & \left. - \: e^{ y_c(\zeta')-y_c(\zeta)} \: \mathrm{erfc}\left( \frac{y_c(\zeta')-y_c(\zeta)+\sigma_y^2(\zeta)+\sigma_y^2(\zeta')}{\sqrt{2 \: \sigma_y^2(\zeta') - 2 \: \sigma_y^2(\zeta)}} \right) \right\} \; .
    \end{split}
\end{equation}

If we assume a constant beam envelope $\sigma_y(\zeta) = \sigma_y = \mathrm{const}$ and replace Eq.~\ref{eq:avedytausol} in Eq.~\ref{eq:avefytau}, we arrive at Eq.~3 of the main text.

\section{Parameters of particle-in-cell simulations}

The 2D Cartesian simulations in Figs. 1a) and 2 of the main text were conducted with a moving window, a grid size $(\Delta \zeta, \Delta y) = (0.08, 0.015)~k_p^{-1}$, a time step $\Delta t = 0.0095~\omega_p^{-1} $, a cold plasma slab extending to $|y| = 1.9~k_p^{-1}$ (with a gap between it and the transverse boundary at $|y| = 2~k_p^{-1}$), and 2 and 4 macroparticles per cell for the plasma and beam electrons, respectively. The boundary conditions were longitudinally open and transversely conducting for electromagnetic fields, and open for macroparticles.

For the 3D simulations in Figs. 3 and 4 of the main text, the grid size is $(\Delta \zeta, \Delta y, \Delta x) = (0.08, 0.008,0.008)~k_p^{-1}$, the time step is $\Delta t = 0.005~\omega_p^{-1} $, and there are two macroparticles per cell for each species. The cold plasma has the same shape as the moving window $140 \times 2 \times 2~k_p^{-1}$, with a  gap of $0.1~k_p^{-1}$ to the open boundaries.

These parameters ensure numerical convergence.

\section{Phase shift measurement method}

\begin{figure}
    \centering
    \includegraphics[width=0.7\textwidth]{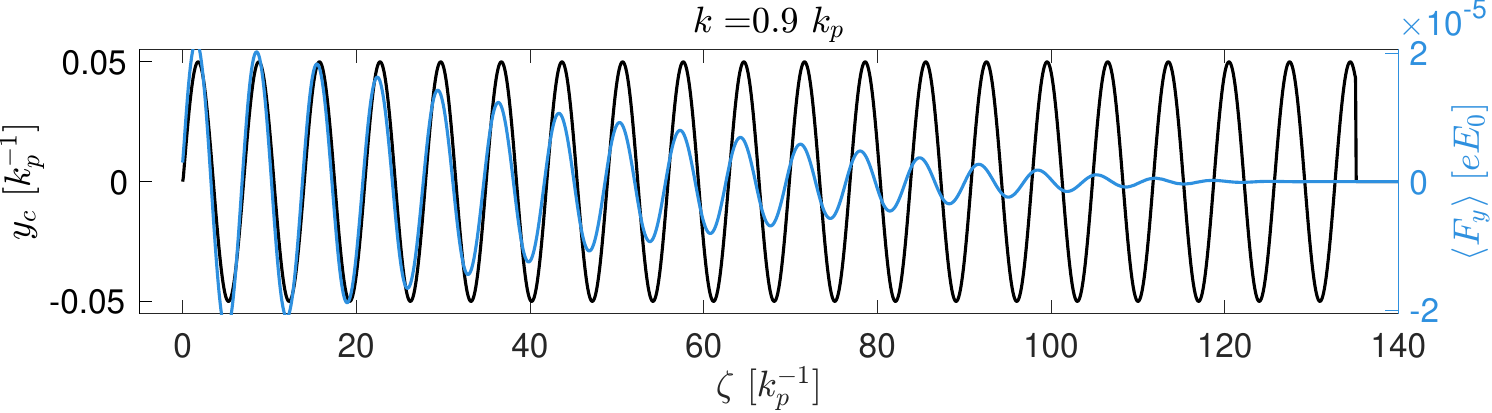}
    \caption{Initial centroid (black line) and initial average transverse force (blue line) from a 2D slab particle-in-cell simulation.}
    \label{fig:curves}
\end{figure}

Our objective is to measure the phase shift between two periodic curves with the same oscillating frequency. We use a cross-correlation of both curves to achieve this, which could be understood as shifting them with respect to each other and finding the amount of shift where their superposition is maximised.

As an example, Fig.~\ref{fig:curves} shows the initial centroid and initial plasma response from a 2D simulation where the wavenumber for the centroid perturbation is $0.9~k_p$. The cross-correlation for these two curves is defined as
\begin{equation}
    \left(y_c \star \left< F_y \right>\right)(\zeta') = \int_{-\infty}^\infty y_c(\zeta) \: \left< F_y \right>(\zeta+\zeta') ~ d\zeta \; ,
\end{equation}
\noindent or, for a finite region $L$,
\begin{equation}
    \left(y_c \star \left< F_y \right>\right)(\zeta') = \int_{-L}^L y_c(\zeta) \: \left< F_y \right>(\zeta+\zeta') ~ d\zeta \; .
\end{equation}

\begin{figure}
    \begin{overpic}[width=0.6\linewidth]{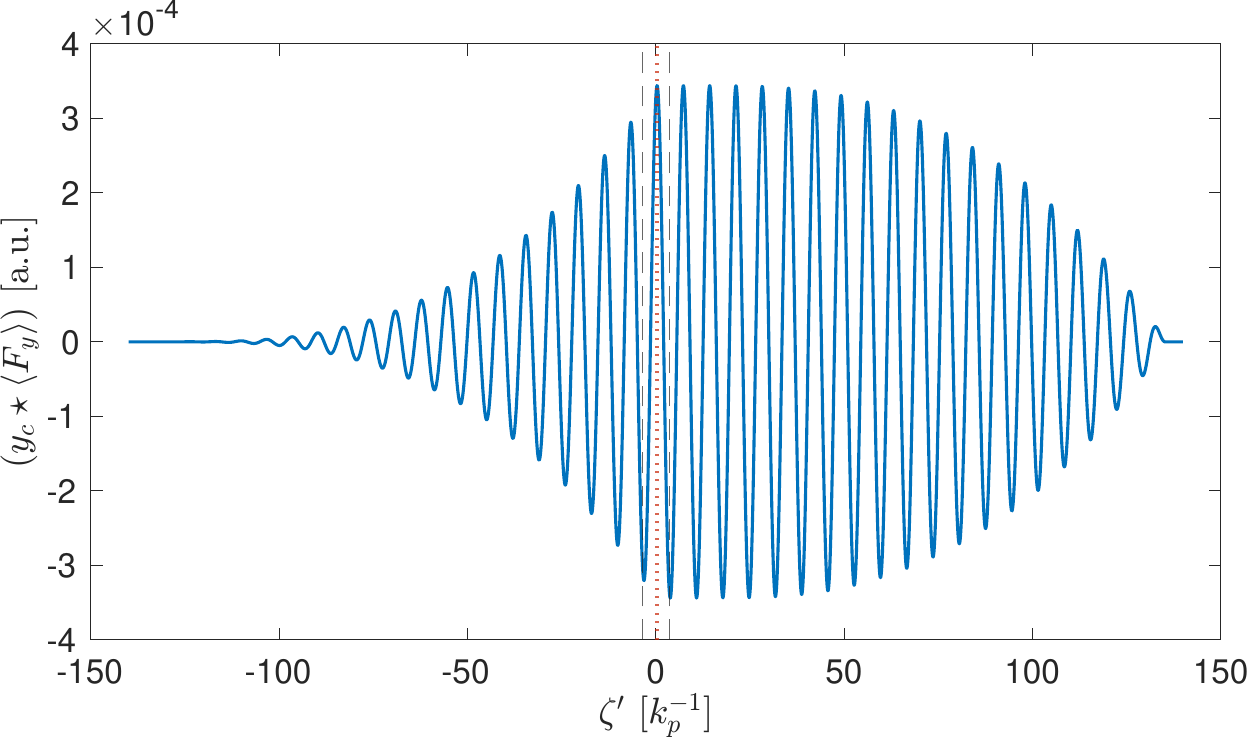}
    \put (12,50) {a)}
    \end{overpic}
    \begin{overpic}[width=0.355\linewidth]{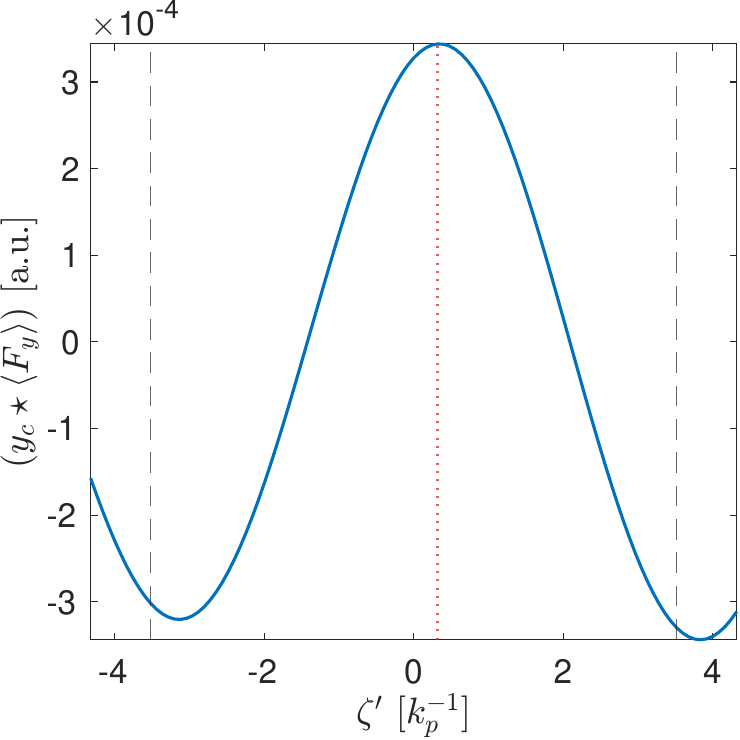}
    \put (25,85) {b)}
    \end{overpic}
 \caption{Cross-correlation of the curves in Fig.~\ref{fig:curves}: a) full data and b) detail. The dashed grey lines indicate the limits $\zeta' = \pm \lambda/2$ and the red dotted line indicates the location of the maximum within this region.}
 \label{fig:xcorr}
\end{figure}

The cross-correlation for the curves in Fig.~\ref{fig:curves} is shown in Fig.~\ref{fig:xcorr}a). We are interested in finding the amount of shift $\zeta'$ within one wavelength $\lambda = 2 \pi / k$ that leads to the most superposition between both curves, i.e. that maximises the cross-correlation. We therefore find the location of the maximum of $\left(y_c \star \left< F_y \right>\right)$ within $\zeta' = \pm \lambda/2$, as illustrated in Fig.~\ref{fig:xcorr}b). The red dotted line denotes the shift $\zeta'_\mathrm{max}$ found in this case. The final phase shift in radians is calculated according to $\Delta \phi = 2 \pi \cdot \zeta'_\mathrm{max}/\lambda$. In Fig.~\ref{fig:xcorr}, for example, $\zeta'_\mathrm{max} = 0.32~k_p^{-1}$ and $\lambda \approx 6.98~k_p^{-1}$, which yields the phase shift $\Delta \phi \approx 0.29~\mathrm{rad}$.

\section{Initial relationship between the centroid and centroid velocity}

In the main text we assume an initial sinusoidal perturbation in the centroid, given by $y_{c0}(\zeta) = \delta_c \sin(\hat{k} \: \zeta)$, where $\delta_c$ is an arbitrary seed amplitude, to study the properties of the hosing instability. Let us now make a reasonably general assumption about the evolution of this initial centroid in the plasma, namely:
\begin{equation}
  y_c(\zeta, z) = A(z)~\sin[\hat{k} \: \zeta - \varphi(z)] \; ,
\end{equation}
where $A(z)$ and $\varphi(z)$ are a time-evolving oscillation amplitude and phase shift, respectively. Figure 1b) of the main text suggests that the centroid amplitude does not change considerably during the early propagation in plasma, hence we may assume $dA/dz \approx 0$ during this phase. The corresponding evolution of the (normalized) centroid velocity, defined as $v_c = v_{c,*}/c = dy_c/dz$, would therefore be given by
\begin{align}
  v_c(\zeta,z) &= - A(z)~\varphi'(z)~\cos[\hat{k} \: \zeta - \varphi(z)] \nonumber \\
  &=  A(z)~\varphi'(z)~\sin[\hat{k} \: \zeta - \varphi(z) - \frac{\pi}{2}] \; ,
\end{align}
where the prime denotes a derivative in $z$. The centroid velocity $v_c \propto \sin[\hat{k} \: \zeta - \varphi(z) - \frac{\pi}{2}]$ and the centroid $y_c \propto \sin[\hat{k} \: \zeta - \varphi(z)]$ are therefore phase-shifted by $\frac{\pi}{2}$.
Since the growth regime is determined by the relative phase shift of $\left< F_y \right>$, this means that the initial plasma response will impact $y_c$ and $v_c$ differently. As $dA/dz$ becomes non-negligible along the propagation, a second term of $v_c$ may become more dominant.

The reasoning above demonstrates that we must alternate between the two non-resonant growth regimes in order for an attempt at mitigation to work, such that we can suppress both a centroid displacement (represented by $y_c$) and transverse momentum (represented by $v_c$).

\bibliography{supmat_references}